# Enhanced thermo-optical response by means of anapole excitation

Javier González-Colsa[1], Juan D. Olarte-Plata[2], Fernando Bresme[2], Pablo Albella[1],*

[1]*Group of Optics, Department of Applied Physics, University of Cantabria, 39005, Santander, Spain.*
[2]*Department of Chemistry, Molecular Sciences Research Hub, Imperial College London, W12 0BZ UK.*

**Corresponding Author**

*pablo.albella@unican.es

**High refractive index dielectric nanostructures offer a versatile platform to control light-matter interaction at the nanoscale as they can easily support electric and magnetic modes with low losses. An additional property that makes them extraordinary is that they can support low radiative modes, so-called anapole modes. In this work, we propose a spectrally tunable anapole nanoheater based on the use of a dielectric anapole resonator able to amplify ten-fold the thermal response of a plasmonic nanoheater. This would allow the use of lower light intensities to achieve striking heating effects. As proof of concept, we perform a detailed study of the thermo-plasmonic response of a gold nanoring used as heating source and a silicon disk, designed to support anapole modes, located in its center acting as anapolar resonator. Furthermore, we utilize the anapole excitation to easily shift the thermal response of these structures from SWIR to the NIR range.**

All-dielectric nanostructures based on high refractive index (HRI) materials offer the possibility to efficiently confine and manipulate light at the nanoscale. [1–3]. The wide variety of possible optical modes across the visible (VIS)[4], near-infrared (NIR)[5] and mid-infrared (MIR)[6] have been exploited in a wide variety of applications[7], such as electric and magnetic hotspot generation[8], strong metal/dielectric coupling[9], Purcell enhancement[10], highly directional scattering[11] or resonant absorption among others. Conversely, plasmonic nanoparticles present high resistive losses enabling a rapid and remarkable temperature increment of their surroundings. This effect is enhanced at the plasmonic resonance which is determined by the free-electron charge-density features (Frölich condition), the particle shape and size. The sudden electron-density oscillations lead to an electric field enhancement that decays rapidly with the distance to the particle surface. These physical effects have been exploited in a wide range of applications such as sensing [12,13], SERS, SEIRA or photoinduced heating [14–19] including photothermal therapies (PTT) [20–22]. However, when the nanostructures become of the order of the excitation wavelength, the description of their electromagnetic response requires three multipolar series: the magnetic, electric, and toroidal [23,24]. A toroidal dipole in combination with an electric one can produce a non-radiating charge current configuration known as dynamic anapole [25–29]. This state appears for a particular wavelength where the fields radiated by the toroidal and electric dipoles cancel each other via destructive interference. An ideal anapole excitation does not emit or absorb, and consequently it cannot be detected in the far field. However, it results in a strong electric field enhancement. This enhancement has been used to develop and improve optical techniques such as Raman scattering[30], refractive index sensing [31,32], narrowband absorption[33] or even optothermal enhancement via lossy dielectrics [34–36]. The study of anapole mode excitations has attracted interest recently, with current efforts directed to boost its efficiency beyond the near-field limit. This enhanced efficiency can be achieved with single all dielectric ring-disk structures. However, to the best of our knowledge previous studies have disregarded the heating potential of metal-dielectric structures assisted by anapole modes, only focusing on the self-induced heating due to the low intrinsic absorption in HRI materials[34], that require excitation power densities that go beyond the conventional ones (up to 24 mW/μm2). In this work, we propose a spectrally tunable anapole nanoheater able to amplify ten-fold the thermal response of a plasmonic nanoheater based on the use of a dielectric anapole resonator. This would allow the use of lower light intensities to achieve striking heating effects. To show this proof of concept, we consider a gold nanoring structure, as heating source. This nanoparticle geometry provides enhanced heating [20]. Then, we locate the HRI dielectric disk, designed to support anapole modes in its center. The thermal amplifying mechanism emerges from the excitation of the anapole mode that develops inside the nanoring which confines the electric field for times longer that it would be possible in absence of the anapole mode [29]. The plasmonic ring surrounding the anapole resonator absorbs this local electric field, consequently suffering a strong amplification of its resistive losses and its conversion into heat. We also present a simple approach to achieve high temperatures in the NIR, a spectral region of particular interest in biomedical applications.

We have used COMSOL Multiphysics to perform all the calculations in combination with numerical FDTD (Finite Difference Time Domain). The heat transport calculations were performed in two steps [37,38]. The first consists on solving the electromagnetic problem to obtain the volumetric distribution of the resistive losses, that is, we solve the Maxwell equations with the RF COMSOL suite. To do so, we have illuminated the hybrid system with linearly polarized light, considering a free tetrahedral mesh with element sizes controlled by the excitation wavelength to guarantee a high element density and reliable curvatures when needed. On the other hand, to consider the heat dissipation, we used a heat flux node across the outer boundaries, considering a heat transfer coefficient, dependent on the geometry and the ambient conditions. We have considered conduction as the main transfer mechanism as we are considering small structures in a stationary fluid. We also neglect the effect of the interfacial thermal conductance since this parameter would increase the inner temperature without changing the external one, thus, heating up the fluid similarly[39]. All the thermal properties involved in this study (density, specific heat and thermal conductivity) were taken from the COMSOL Multiphysics material database. Figure 1a depicts a schematic of the anapole excitation produced within a silicon disk of 340 nm radius when illuminated at a wavelength $\lambda \approx 1300$ nm [40]. It can be observed how the electric field lines follow the toroidal geometry shown in the upper inset. The electric field circulation produces the distinct anapole pattern (shown in the lower inset) that can be characterized by a sudden drop in the scattering spectra accompanied by a lack of absorption.

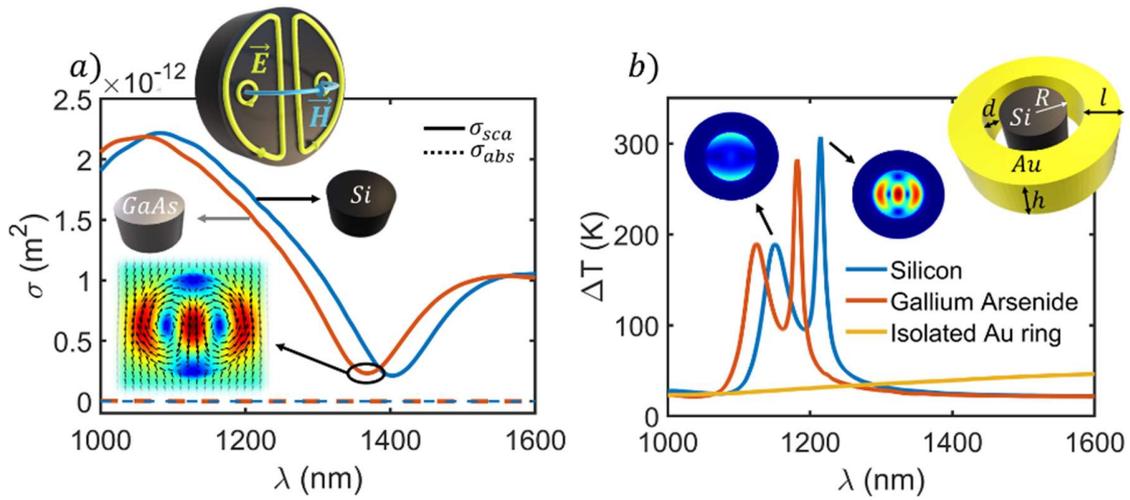

**Figure 1:** a). Scattering (solid line) and absorption (dotted line) cross sections for silicon (blue) and gallium arsenide (orange) disks with $R$=340 nm and $h$=155 nm [40]. Anapole mode electric field distribution (schematically represented in the inset). b) Comparison of the spectral temperature increment distribution for the isolated gold ring and the optimized anapole generators (outlined in the upper inset where $d$ dielectric-metallic distance, $R$ the disk radius, $l$ the ring length and $h$ the height) defined by $d$=0 nm, $R$=340 nm, $l$=290 nm and $h$=155 nm. The incident power density is fixed at 0.1 mW/μm$^2$.

The HRI disk scattering and absorption cross section spectra are shown in Figure 1a. Here, a spectral drop in the scattering cross section can be clearly seen for both materials together with null in the absorption cross section. This fact, together with the electric field spatial distribution (shown in Figure S1) evidences the excitation of a strong anapole mode. Furthermore, their cross sections reveal that the anapole mode spectrally differs for Si and GaAs disks, being the response on the case of silicon redshifted with respect to the gallium arsenide. This is a consequence of their slightly different optical constants. Besides that, materials like Si or GaAs have a very low imaginary part of the refractive index in the near and MID infrared translating into a negligible absorption. Figure 1b illustrates the analyzed hybrid nanostructure built by a metallic nanoring (made of gold) and the HRI disks shown in figure 1a (two different materials are explored, Si and GaAs). The optimal structure is found by a careful electromagnetic optimization of the ring parameters leading to ring defined by $l$ = 290 nm, $d$= 0 nm and $h$ = 155 nm (see Figure S2-3 of the supplementary document for more information). Although fabrication may be challenging, complex structures can be made by high-resolution techniques such as e-beam lithography[41], sequential lithographic deposition/etchings for each material [42] and Critical Energy Electron Beam Lithography[43]. The results show that the ring parameters mainly influence the ability of the nanomaterial to reach high temperatures, since the resistive loss is enhanced as gold approaches the HRI resonator. Thus, the optimal heating response is reached for full contact between both materials $d = 0\ nm$. In Figure 1b the thermal spectral response of the optimal hybrid system is also compared with an isolated ring showing that the ring thermal response is about 10 times lower than the hybrid one. It can be explained by the anapole excitation which boosts the ring electromagnetic response acting as the electromagnetic resonator. This enhanced electric field induces a more intense Joule effect leading to an increment of the resistive losses. In particular, the temperature increment is amplified about 10 times for the Si and more than 9 for GaAs resonators for an intensity of 0.1 mW/μm². The low absorption in the disk resonator is a key parameter, since highly absorptive materials reduce the anapole effectivity affecting the resistive losses in the gold ring. This translates in a shorter response and in a weaker heating. We have analyzed the dependence of the nanomaterial heating with its geometry, and used this information to guide the design of optimum anapole nanomaterials, targeting their heating performance.

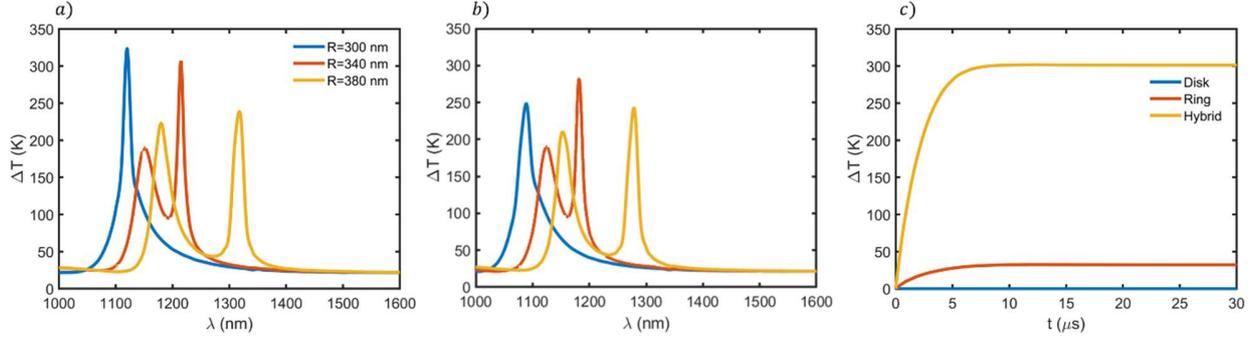

**Figure 2:** Spectral thermal response of the optimal hybrid ring/disk resonator calculated for different disk radii for silicon (a) and gallium arsenide (b). The ring length and height are fixed to $d$=0 nm and $h$=155 nm respectively. c) Transient thermal state for the isolated silicon disk (blue), ring (red) and the hybrid structure (yellow).

As shown in Figure 2a)-b), the disk radius gives rise to remarkable changes in the thermal spectrum of both, Si and GaAs resonators. The anapole mode is redshifted for larger radii, as expected. This supports the notion that a hybrid disk/ring resonator can be easily tuned within the NIR, where a ring-shaped gold nanostructure of the same size is unable to resonate[44,45]. Notice that the double spectral resonances merge as radius approaches 300 nm. Figure 2c) shows the transient thermal state calculated within the structure for the isolated ring and disk compared with the hybrid unit cell. The temporal evolution is negligible in the case of the isolated silicon disk. It can be understood since silicon does not present losses within the NIR. Despite the ring-shaped and hybrid structures present remarkable differences in temperature, it can be seen that they reach the stationary state for similar periods of around 5 microseconds approximately. We have presented above a comprehensive study of the hybrid disk-ring performance as a heating unit. We now explore its behavior when a substrate is present. To do so, we consider the configuration defined by $R$=340 nm, $l$=290 nm, $d$=0 nm and $h$=155 nm. Commonly used substrates with contrasting thermal properties have been considered to demonstrate the huge impact that the thermal conductivity has on the photothermal capabilities of the proposed structure. We have selected alumina, silica and PDMS as examples of high, middle and low thermal conductivities with 24 W/mK, 1.3 W/mK and 0.15 W/mK respectively. In Figure 3a, the scheme of the hybrid structure on a substrate of thermal conductivity $k$ can be seen. We consider a power density of 0.1 mW/μm² and linearly polarized light to excite the hybrid structure at normal incidence. In Figure 3b, the spectral thermal response of the three substrates is obtained. All curves show the spectral resonance at similar wavelengths as the considered materials have similar optical properties. However, they present radically opposed thermal responses as a consequence of their thermal conductivity contrast. A well-defined trend can be extracted from the maximum temperature increment.

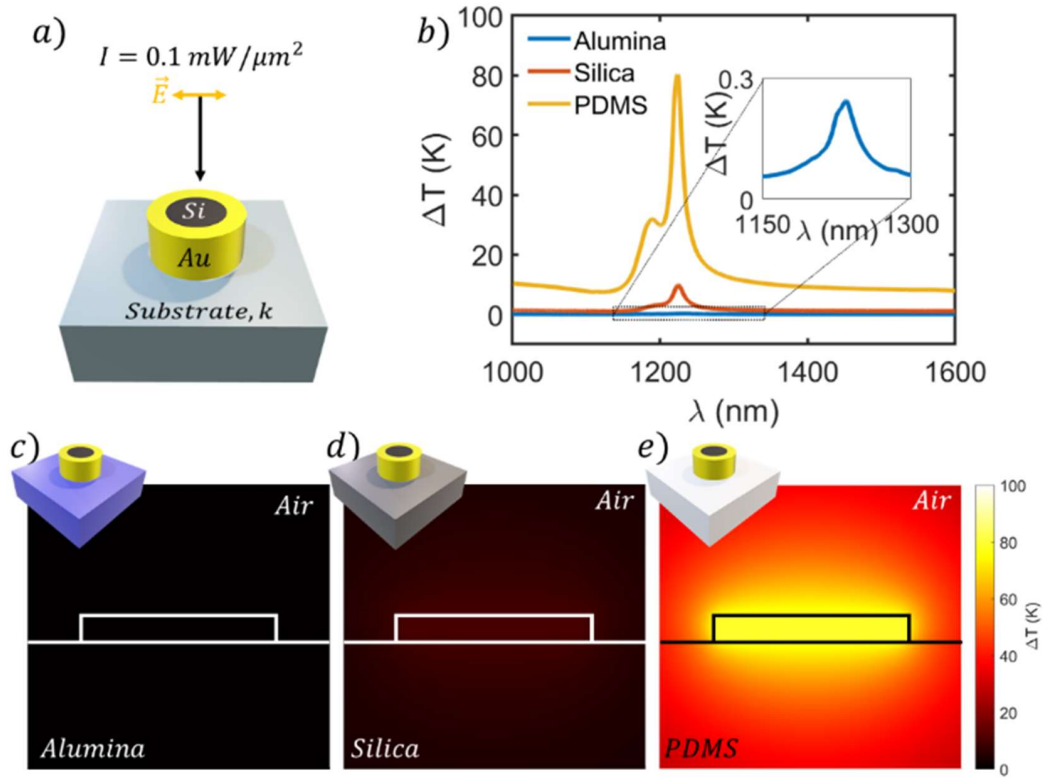

**Figure 3.** Thermal behavior of the proposed structure on different conductive substrates. a) Scheme of the hybrid structure on a substrate of thermal conductivity $k$ illuminated at normal incidence with a power density of 0.1 mW/μm$^2$. b) Temperature increment spectra for the three considered materials, i.e., alumina (blue), silica (red) and PDMS (yellow). Temperature increment spatial distribution for a substrate of alumina (d), silica (e) and PDMS (f).

As the substrate thermal conductivity grows, the temperature decreases. The temperature is also smaller with respect to the isolated hybrid structure as air has a much lower thermal conductivity. Thus, the PDMS substrate shows the best performance allowing for temperature increments of about 80 K. Following Figures c)-e), the alumina substrate features a negligible heating effect and the silica substrate is a 90% less effective than the PDMS one. This result can be understood by considering the high conductivity of substrates, and how the heat flows preferentially through the higher thermal conductivity materials, so that the stationary temperature increment is reduced significantly. Conversely, low thermal conductivity materials inhibit heat flow leading to a higher increase of the nanomaterial temperature in the stationary state.

In summary, we have demonstrated that anapole excitation can serve as an easy mechanism to boost the heating effect in ring-shaped gold structures. Additionally, their thermal response can be tuned to the NIR, where they are not able to resonate a priori. Its thermal performance was analyzed by placing them on different thermal conductivity substrates (alumina, silica and PDMD), obtaining temperature increments ten-fold that of single gold rings. We believe that the reliable implementation of our proof of concept can motive the development of novel strategies to reach efficient nanoheating structures and temperature-controlled platforms that will be of general interest to the thermoplasmonic community.

ASSOCIATED CONTENT

**Supporting Information**. Additional computational details, materials, descriptions, including the optimization of the proposed anopole nanoheater.

AUTHOR INFORMATION

The authors declare no competing financial interests.

ACKNOWLEDGMENT

Authors would like to thank Prof C. R. Crick for the interesting and valuable discussions. We gratefully acknowledge financial support from Spanish national project INMUNOTERMO (No. PGC2018-096649-B-I), the UK Leverhulme Turst (Grant No. RPG-2018-384). J. G-C. thanks the Ministry of science of Spain for his FPI grant and P.A. acknowledges funding for a Ramon y Cajal Fellowship (Grant No. RYC-2016-20831).

# Supplementary document

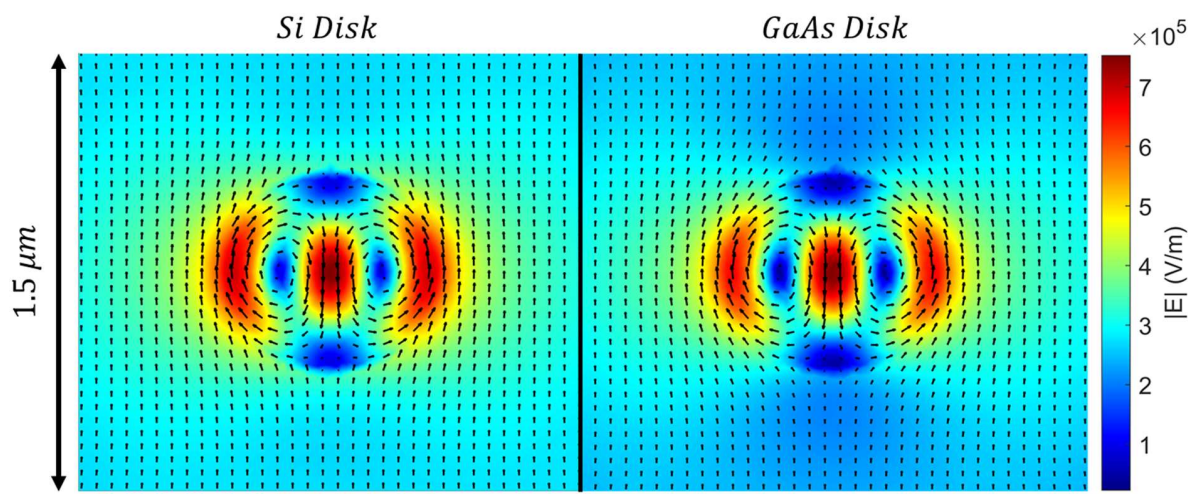

**Figure S1.** Electric field distributions of the resultant anapole mode excitation in silicon and gallium arsenide disks at $\lambda = 1400$ nm and $\lambda = 1366$ nm. The disk is immersed in air and has a height $h = 155$ nm and a radius $R = 340$ nm.

Figure S1 shows the electric field distribution of the anapole modes induced within a silicon and gallium arsenide disks excited at a wavelength of 1400 nm and 1366 nm respectively. The resonance wavelength is slightly different due to the refractive index contrast between both materials. Thus, both systems show similar electric field pattern and lines that match with the expected ones for an anapole mode excitation.

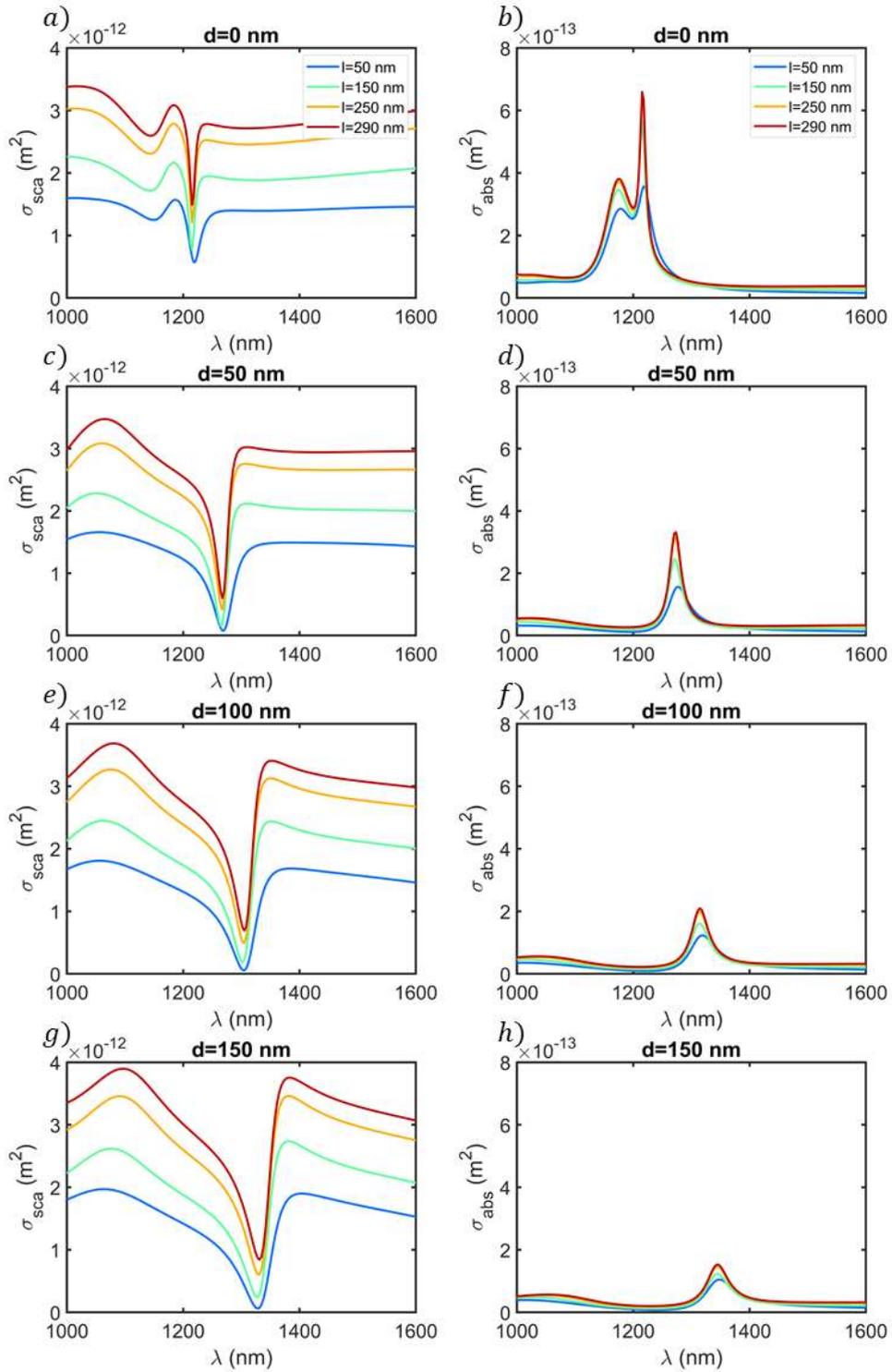

Figure S2. Scattering and absorption cross section of a hybrid silicon-gold disk/ring structure for a set of ring lengths ($l$) from 50 to 290 nm for $d = 0\ nm$ (a, b), $d = 50\ nm$ (c, d), $d = 100\ nm$ (e, f) and $d = 150\ nm$ (g, h), respectively.

In Figure S2 the behavior of the hybrid silicon/gold structure absorption and scattering cross section can be seen. In general, the scattering (Figure a, c, e, g) increases for thicker gold rings as expected, however, the spectral resonance remains constant. The $l$ parameter also boosts the absorption due to the gold volume increment. Focusing on the absorption (Figure b, d, f, h), the most influential geometrical parameter is the metal-dielectric distance. As $d$ goes to zero the absorption increases which is translated in a thermal performance improvement. This can be explained in terms of the anapole resonance. When the anapole is generated inside the disk, an evanescent field is created in the near field so that, the nearer the gold is to the dielectric interface, the stronger the evanescent field influence becomes. This leads to a resistive loss enhancement in the metal. It is also noticed a moderate blueshift because of the ring main radius reduction. Thus, in sight of Figure S2, the optimal disk-ring configuration in terms of light to heat conversion is $d = 0\ nm$ and $l = 290\ nm$.

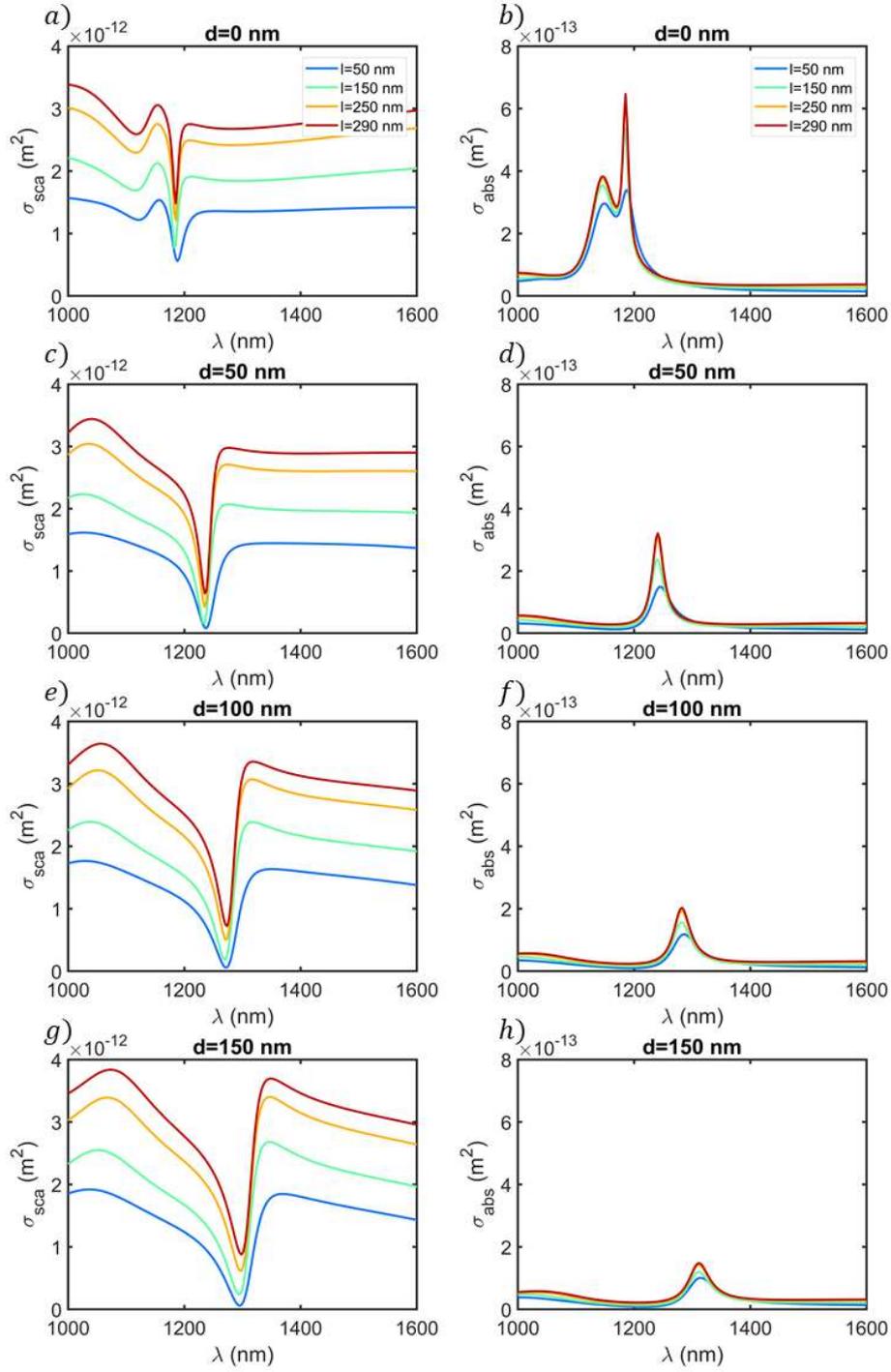

**Figure S3**. Scattering and absorption cross section of a hybrid gallium arsenide-gold disk/ring structure for a set of ring lengths ($l$) from 50 to 290 nm for $d = 0\ nm$ (a, b), $d = 50\ nm$ (c, d), $d = 100\ nm$ (e, f) and $d = 150\ nm$ (g, h) respectively.

In Figure S3 the behavior of the hybrid gallium arsenide/gold structure absorption and scattering cross section can be seen. It shows a similar behavior to the Figure S2 so that the same conclusions are extracted. Thus, the optimal disk-ring configuration in terms of light to heat conversion is also $d = 0\ nm$ and $l = 290\ nm$.